\documentclass[prl,aps,twocolumn,showpacs]{revtex4}

\usepackage{graphicx}

\begin{document}

  \title{Charge Order and the Origin of Giant
    Magnetocapacitance in LuFe$_2$O$_4$}

 \author{H. J. Xiang}
 \affiliation{Department of Chemistry, North Carolina State University, Raleigh,
 North Carolina 27695-8204}

 \author{M.-H. Whangbo}
 \thanks{Corresponding author. E-mail: mike\_whangbo@ncsu.edu}

 \affiliation{Department of Chemistry, North Carolina State University, Raleigh,
 North Carolina 27695-8204}

\date{\today}

\begin{abstract}
  The nature of the charge order in the charge frustrated compound
LuFe$_2$O$_4$ and its effect on magnetocapacitance were examined on the basis of
first-principles electronic structure calculations and Monte Carlo
simulations of electrostatic energy. Our work shows that 
two different types of charge order of almost equal stability
(i.e., $\sqrt{3}\times \sqrt{3}$ and chain types) occur in the
Fe$_2$O$_4$ layers of LuFe$_2$O$_4$, and that the
ground state of LuFe$_2$O$_4$ has a ferrielectric arrangement of the
Fe$_2$O$_4$ layers with $\sqrt{3}\times \sqrt{3}$ charge order. The giant
magnetocapacitance effect of LuFe$_2$O$_4$ at room temperature is 
accounted for in terms of charge fluctuations arising from the interconversion
between the two types of
charge order, that becomes hindered by an applied magnetic field. 
\end{abstract}

\pacs{71.20.-b,71.45.Lr,77.80.-e,77.84.-s}
%71.20.-b Electron density of states and band structure of crystalline
%solids
%71.45.Lr Charge-density-wave systems (see also 75.30.Fv Spin-density waves)
%77.80.-e Ferroelectricity and antiferroelectricity
%77.84.-s Dielectric, piezoelectric, ferroelectric, and antiferroelectric materials

\maketitle
Ferroelectric (FE) oxides are essential components in a large number of
applications \cite{Auciello1998}.
In traditional FE materials like
BaTiO$_3$, the ferroelectricity is driven by the hybridization of the
empty d orbitals of Ti$^{4+}$ with the occupied p orbitals of the
oxygen anions \cite{Cohen1992}. Recently, a
mixed-valence compound LuFe$_2$O$_4$ with the average valence Fe$^{2.5+}$ was
found to exhibit ferroelectricity associated with the charge order
(CO) leading to Fe$^{2+}$ and Fe$^{3+}$ ions \cite{Ikeda2005}. Subramanian {\it
  et al.} reported that at room temperature (RT) the dielectric constant of
LuFe$_2$O$_4$ decreases sharply when a small magnetic field  is applied
\cite{Subramanian2006}. This suggests a strong coupling between spin
moment and electric dipole at RT, and hence potential
applications of LuFe$_2$O$_4$ in which the charge and spin degrees of
freedom of electrons can be controlled.

At RT LuFe$_2$O$_4$ has a hexagonal layered structure (space
group R\=3m, a = 3.44 \AA, and c = 25.28 \AA)  in which all Fe sites
are crystallographically  equivalent \cite{Isobe}. LuFe$_2$O$_4$ is an
insulator \cite{Tanaka}, and undergoes a three-dimensional CO
below 330 K \cite{Isobe,LuFe2O4,RFe2O4} as well as a two-dimensional
ferrimagnetic order below 240 K \cite{RFe2O4}. In LuFe$_2$O$_4$,
layers 
of composition Fe$_2$O$_4$ alternate with layers of Lu$^{3+}$ ions, and there are
three Fe$_2$O$_4$ layers per unit cell (Fig. 1(a)). Each Fe$_2$O$_4$ layer (referred
to as the W-layer) is made up of two triangular sheets of
corner-sharing FeO$_5$ trigonal bipyramids (Fig. 1(b)-(d)).

\begin{figure}[!hbp]
  \includegraphics[width=6.0cm]{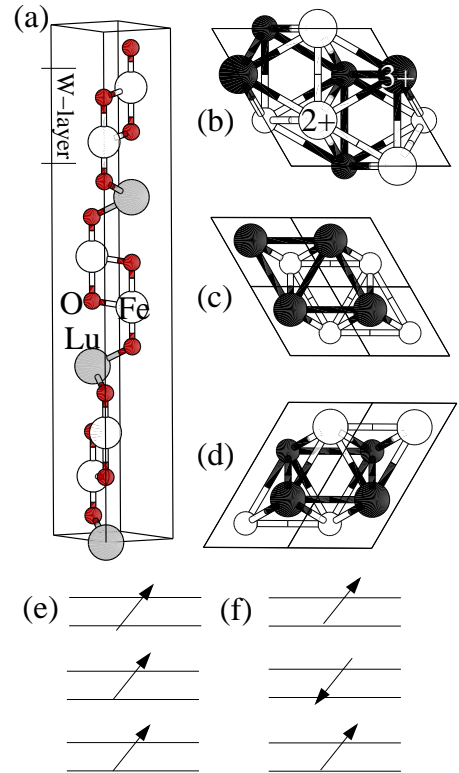}
  \caption{(Color online) (a) Hexagonal unit cell structure of
  LuFe$_2$O$_4$, where grey, white and red circles represent the Lu,
  Fe and O atoms, respectively.
  Schematic representations of the (b) CO-I, (c) CO-II and
  (d) CO-III structures, where the small and large circles refer to
  the Fe ions in the upper and lower triangular sheets of a W-layer,
  respectively. The Fe$^{2+}$ and Fe$^{3+}$ ions are represented by
  empty and filled circles, respectively. 
  Schematic representations of the (e) FE and (f) FIE
  arrangements of the dipoles of the W-layers in the CO-I structure.}
  \label{fig1}
\end{figure}

LuFe$_2$O$_4$ exhibits apparently puzzling physical properties, and the
nature of its CO is not unequivocal.
It is generally believed that
lattice distortions accompany a CO \cite{Wright2001}. 
Thus, it is unclear why LuFe$_2$O$_4$ is insulating above RT
despite that  LuFe$_2$O$_4$  adopts the structure in which all Fe sites are
equivalent in this temperature region.
The giant magnetocapacitance effect found for LuFe$_2$O$_4$ can be
understood by supposing that the charge fluctuation (CF) of LuFe$_2$O$_4$ arising
from its charge frustration is sharply reduced by an external
magnetic field. However, it is unclear by what mechanism this happens. As  
to the nature of the CO in LuFe$_2$O$_4$, there
is a controversy. 
By analogy with the stable long-range order of Ising spins in a
triangular lattice antiferromagnet (TLA), Yamada {\it et al.} proposed
a model CO structure in which the two triangular sheets of a W-layer do not have
the same number of Fe$^{2+}$ and Fe$^{3+}$ ions, i.e., [Fe$^{2+}$]:[Fe$^{3+}$] = 1:2 in
one triangular sheet, and 2:1 in another triangular sheet (i.e., the
CO-I structure in Fig. 1(b)) \cite{Yamada1997}. 
The  $\sqrt{3}\times \sqrt{3}$  superstructure of this CO model is
compatible with the experimental observation. 
Subramanian {\it et al.} proposed a different CO structure in which
one triangular sheet of a W-layer has only Fe$^{2+}$ ions and the
other triangular sheet
has only Fe$^{3+}$ ions (i.e., the CO-II structure in Fig. 1(c))
\cite{Subramanian2006}, on the basis of the fact that the closest
Fe-Fe distances in  LuFe$_2$O$_4$ occur between adjacent triangular
sheets of a W-layer rather than within each triangular sheet.

In this Letter, we probe the nature of the CO and the origin of the magnetocapacitance
in LuFe$_2$O$_4$ on the basis of first principles electronic structure
calculations and Monte Carlo (MC) simulations of electrostatic
interactions. Our study shows that the ground state of LuFe$_2$O$_4$ has the
CO-I structure, the CO-II structure is unstable, and another CO
structure  different from CO-I and CO-II  is very close in
energy to the CO-I structure. The presence of two different CO
structures close in energy is found crucial  for the magnetocapacitance effect
of LuFe$_2$O$_4$. 

Our 
spin-polarized density functional theory 
calculations were performed on the basis of the frozen-core projector
augmented wave method \cite{PAW} encoded in the Vienna {\it ab initio}
simulation package \cite{VASP} using the generalized-gradient
approximation (GGA) \cite{Perdew1996}
and the plane-wave cut-off energy of 400 eV. 
To properly describe the strong electron correlation in the 3d
transition-metal oxide, the GGA plus on-site repulsion U method
(GGA+U) was employed \cite{Dudarev1998} with the effective $U$ value ($U_{eff} =
U -J $ ) of 4.61 eV. 
Calculations with various $U_{eff}$ values show that our main results
remain valid when $U_{eff}$ is varied between $\sim$3.6 and $\sim$5.6 eV. In the
following we report only those results based on $U_{eff} = 4.61$ eV.
To simplify our discussion and reduce the
computational task, we explore the CO structures of LuFe$_2$O$_4$ under the
assumption that the spins of LuFe$_2$O$_4$  have a ferromagnetic (FM)
ordering. This assumption is reasonable because the energy scale
associated with different spin arrangements is much smaller than that
associated with different CO's. An additional restriction of our
calculations is the use of the experimental lattice constants. Our
full geometry optimization of LuFe$_2$O$_4$ with GGA calculations leads to
the lattice constants that are very close to the experimental values.

The GGA+U calculation for the non-CO state of LuFe$_2$O$_4$ 
with the room-temperature crystal structure and hence the hexagonal symmetry
shows that all Fe ions are in the high spin
state. 
The plots of density of states (DOS) calculated for the non-CO state
(Fig. 2(a)) predict a metallic behavior for LuFe$_2$O$_4$, in disagreement with
experiment. This is not surprising 
because the Fe
ions are in the valence state of $2.5+$ in the absence of CO. We found
it impossible to produce an insulating band gap for any CO structure
as long as the structure keeps the hexagonal symmetry.
For a transition metal ion in a trigonal bipyramidal crystal field,
the five $d$ states are split in three groups, i.e., \{$d_{xy}$, $d_{x^2-y^2}$\}, \{$d_{xz}$,
  $d_{yz}$\} and \{$d_{z^2}$\}. The partial DOS plots shown in Fig. 2(b) indicates that the
down-spin bands leading to a metallic character arise from the $d_{xy}$ and
$d_{x^2-y^2}$ orbitals, i.e., the \{$d_{xy}$, $d_{x^2-y^2}$\} states have a lower energy
than do the \{$d_{xz}$,  $d_{yz}$\} states for the FeO$_5$ trigonal bipyramids in
LuFe$_2$O$_4$.

\begin{figure}[!hbp]
  \includegraphics[width=7.0cm]{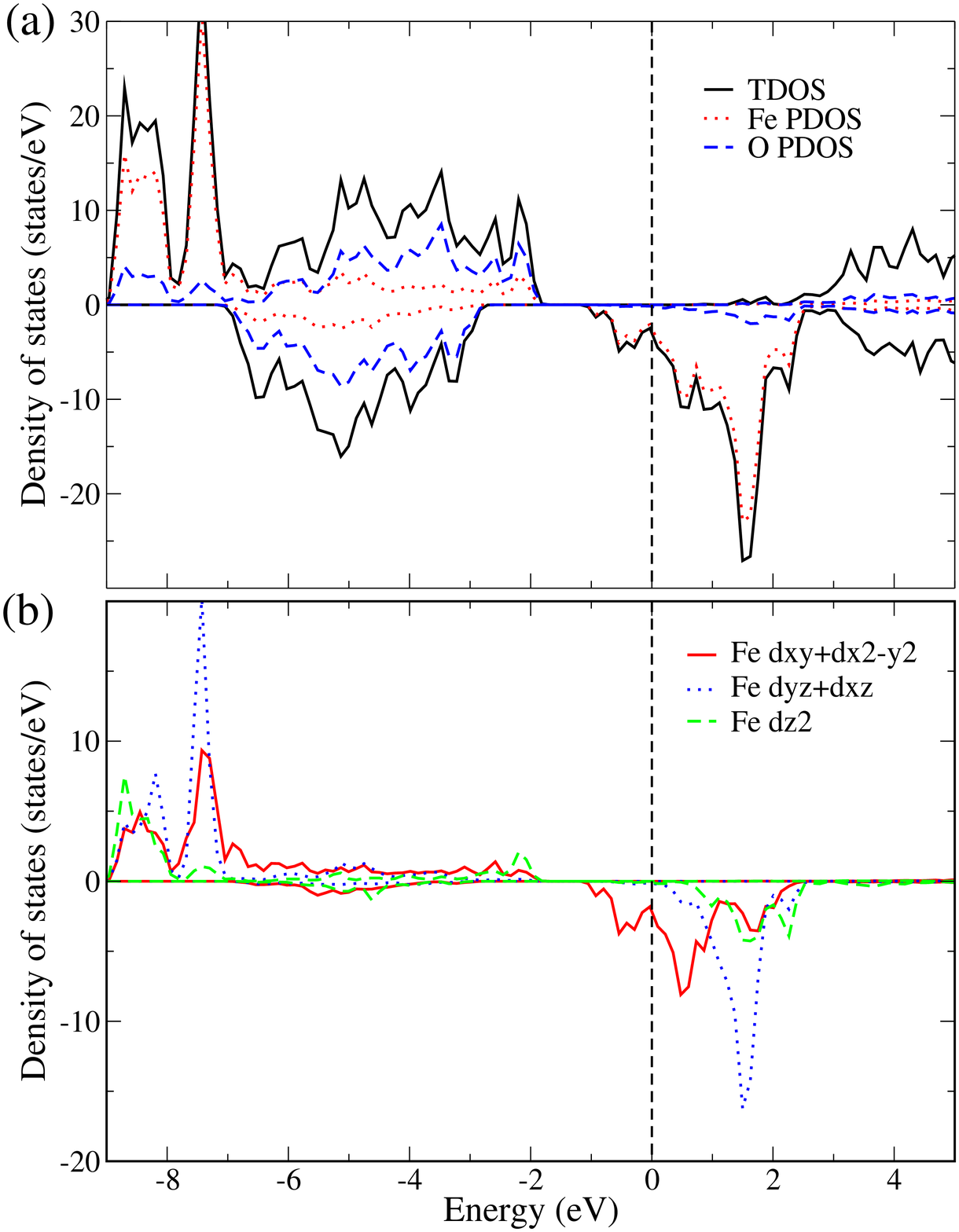}
  \caption{(Color online) DOS of LuFe2O4 calculated for
    the RT structure without CO. The DOS was calculated
    with 0.1 eV broadening.} 
  \label{fig2}
\end{figure}

\begin{figure}[!hbp]
  \includegraphics[width=7.0cm]{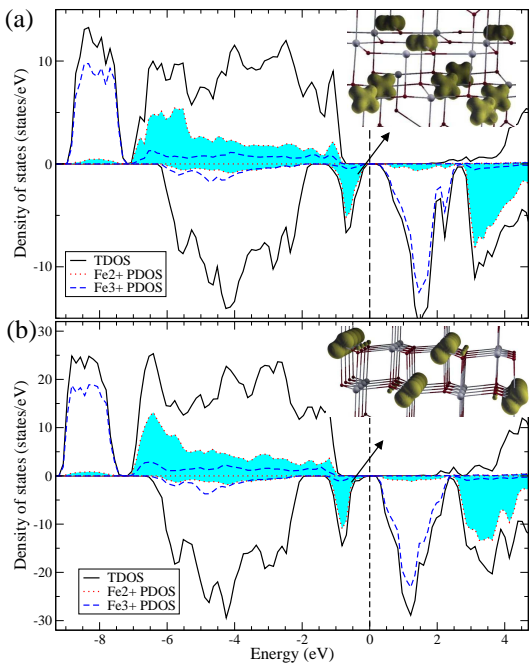}
  \caption{(Color online) 
    DOS of LuFe$_2$O$_4$ calculated for (a) the relaxed CO-I and (b) the relaxed
    CO-III structures. The blue regions highlight the PDOS of the
    Fe$^{2+}$ ion. 
    The inset shows the charge density plot calculated
    for the occupied down-spin d-states. The DOS was calculated with 0.1
    eV broadening.}
  \label{fig3}
\end{figure}

To search for possible low-energy CO patterns, we resort to the
classical MC simulation method by considering only the intersite
Coulomb repulsion. This simplified
approach is reasonable for the purpose of finding stable CO
patterns. As the energy reference of this calculation, each Fe$^{2.5+}$
site of the non-CO state is assumed to carry zero charge, so that,
after a CO takes place, Fe$^{3+}$ and Fe$^{2+}$ sites will carry $0.5\delta$
($0<\delta<1$) and $-0.5\delta$ charges, respectively ($\delta$ denotes the
degree of the charge transfer). 
For simplicity, we assume that the charge on oxygen is independent of
$\delta$. 
We perform MC simulations using a
$2\sqrt{3}\times 2\sqrt{3}$ supercell with periodic boundary
conditions and evaluate the long-range Coulomb interaction using the
Ewald sum method. 
There occur two different low-energy CO patterns. The most
stable one is a chain-like CO structure (hereafter the CO-III
structure) with energy $-1.601 \delta^2$ eV per formula unit (FU), in
which chains of Fe$^{2+}$ ions alternate with those of Fe$^{3+}$ ions in each
triangular sheet (Fig. 1(d)). 
Given the trigonal symmetry of each W-layer, there are three
different, but equivalent, ways of choosing the chain direction in
each W-layer. When the three different chain orientations occur
randomly, the overall superlattice 
diffraction pattern will exhibit a $\sqrt{3}\times \sqrt{3}$
structure. 
The CO-I structure proposed by Yamada {\it
  et al.} is found to have a slightly higher energy, i.e.,
$-1.379 \delta^2$ eV/FU. (The stability of this CO cannot be deduced
by analogy with the $\sqrt{3}\times \sqrt{3}$  spin configuration of
Ising spins that occurs in a TLA when the nearest-neighbor (NN) spin
exchange is antiferromagnetic and the next-nearest-neighbor (NNN) spin
exchange is FM. This spin configuration becomes the ground
state only when the NNN interactions are
stabilizing. In the CO-I structure, the NNN
interactions are destabilizing due to Coulomb repulsion.)
Our MC simulations show that the CO-II structure proposed by
Subramanian {\it et al.} is found to be highly unstable, namely, its
energy is even higher than the non-CO structure by $0.496 \delta^2$
eV/FU. The instability of the CO-II structure is due to the large
positive electrostatic energy between the ions of the same charge in
each triangular sheet.  We also consider MC simulations using another
model that takes into consideration only the NN
electrostatic interactions within each triangular sheet and between
adjacent triangular sheets of every W-layer. Qualitatively, this model
leads the same results as described above. 

To more accurately probe the stabilities of the CO-I, CO-II and CO-III
structures with respect to the non-CO structure, we carry
out GGA+U
calculations for LuFe$_2$O$_4$ on the basis of its crystal structure
determined at RT. Our GGA+U calculations for the CO-II structure,
with an initial guess of the charge density expected for it, always
converged to the non-CO structure. Therefore, it is concluded that the
CO-II structure is not stable, as found from our MC simulation. Since
there are three W-layers per unit cell and since each W-layer has a
nonzero dipole moment in the CO-I structure, there are two different
arrangements of the W-layers in the CO-I structure, i.e., the
FE arrangement (Fig. 1(e)) and the ferrielectric (FIE) arrangement
(Fig. 1(f)). For the FE CO-I and the CO-III structures, our GGA+U
calculations were carried out in two steps. In the first step, the
FE CO-I or the CO-III structure was introduced in LuFe$_2$O$_4$
without allowing the crystal structure to relax. 
Such a CO induced solely by electrostatic interactions was first
proposed by Attfield {\it et al.} \cite{Attfield} and later computationally
realized by Leonov {\it et al.} \cite{Leonov} in their studies of the CO
phenomenon in Fe$_2$OBO$_3$. 
In the second step,
the crystal structure with the FE CO-I or the CO-III structure
was completely optimized. Thus, the energy gain $\Delta E$ of the
FE CO-I or the CO-III structure relative to the non-CO structure
is written as $\Delta E = \Delta E_{1}  + \Delta E_{2}$, where
$\Delta E_{1}$  and  $\Delta E_{2}$ are the energy gains obtained in the first
and second steps, respectively. $\Delta E_{1}$ arises from
electrostatic interactions, and $\Delta E_{2}$ from the geometry
relaxation. 

Our GGA+U calculations of the first step show that the
FE CO-I and the CO-III structures are stable in the absence of
geometry relaxation \cite{largeU} (with  $\Delta E_{1}$ $=$ 239 and 219
meV/FU, respectively), and are an insulator (with band gap of 0.18
and 0.15 eV, respectively).
The forces acting on the atoms calculated for both CO
structures with no geometry relaxation are large hence indicating
instability of the ``frozen'' structure. Our GGA+U calculations of the
second step show that the $\Delta E_{2}$ values of the FE
CO-I and the CO-III structures are 163 and 165 meV/FU,
respectively. Thus, after structural relaxation, the FE
CO-I structure remains only slightly more stable than the CO-III
structure (by 19 meV/FU). Another important finding of our
calculations is that the energy gain resulting solely from
electrostatic interactions is greater than that from geometry relaxation
(i.e., $\Delta E_{1}  > \Delta E_{2}$), in contrast to the case of 
Fe$_2$OBO$_3$, for which our calculations showed that  
$\Delta E_{1}  \ll \Delta E_{2}$
(i.e., 77 vs. 272 meV) \cite{To_be_published}.
Bond-valence-sum calculations \cite{BVS} for the optimized
structures of the FE CO-I and the CO-III structures show that
the valence states of the ``Fe$^{2+}$'' and ``Fe$^{3+}$'' sites are close to the
nominal +2 and +3, respectively. Both CO structures have a larger band
gap after geometry optimization (i.e., 0.70 and 0.54 eV for the
FE CO-I and the CO-III structures, respectively). The DOS plots calculated
for the FE CO-I and the CO-III structures with the relaxed
structures are presented in Fig. 3. All the occupied up-spin 3d bands
of the Fe$^{2+}$ ions lead to a spherical charge distribution, and so do
those of the Fe$^{3+}$ ions. It is the occupied down-spin 3d bands of the
Fe$^{2+}$ ions that provide an anisotropic charge distribution arising
from the Fe 3d orbitals and hence
information about the orbital order (OO). This is shown in the insets
of Fig. 3, which reveal that the two triangular sheets of a W-layer
have an identical OO in the CO-III structure, but different OO's in
the FE CO-I structure. 
In the FE CO-I structure, the orbitals of the \{$d_{xy}$, $d_{x^2-y^2}$\}
type are involved in the OO of the triangular sheet of containing
fewer Fe$^{2+}$ ions than Fe$^{3+}$ ions, but those of the \{$d_{xy}$, $d_{x^2-y^2}$\} and
\{$d_{xz}$, $d_{yz}$\} types in the other triangular sheet. In the CO-III
structure, the orbitals of the \{$d_{xy}$, $d_{x^2-y^2}$\}
and \{$d_{xz}$, $d_{yz}$\} types are
involved in the OO of both triangular sheets of a W-layer. 
The OO's found for the unrelaxed
structure are quite similar to those found for the optimized
structures described above.

Experimentally, a large spontaneous
electric polarization (EP) was found in LuFe$_2$O$_4$ \cite{Ikeda2005}. Of the
two stable CO structures, CO-I and CO-III, only the CO-I structure has
a nonzero EP.
For the calculations of the spontaneous
EP of the FE CO-I structure, we use the
Berry phase method \cite{Berry}. 
The calculated EP should be compared with the
experimental one measured at a low temperature where there is no
CF.
The spontaneous EP
along the c-axis is calculated to be 52.7 $\mu C/cm^2$, which is much
greater than the experimental value of about 25 $\mu C/cm^2$ measured
at 77 K \cite{Ikeda2005}. To resolve this discrepancy, we consider the
FIE CO-I structure (Fig. 1(f)). Our GGA+U calculations with
complete geometry optimization show that the FIE CO-I
structure is more stable than the FE CO-I structure by 12
meV/FU. The spontaneous EP along the c-axis
calculated for the
FIE CO-I structure is 26.3 $\mu C/cm^2$, in
excellent agreement with experiment. Thus, from the viewpoint of the
total energy and the spontaneous EP, it is
concluded that LuFe$_2$O$_4$ has the FIE CO-I
structure. Note that the spontaneous EP for the
FIE CO-I structure is smaller than that of FE
CO-I structure by a factor of approximately two, instead of three
expected from Fig. 1(e),(f). This is due to the fact the two crystal
structures, which are each separately optimized, are slightly
different.

Now we turn attention to the probable origin of the giant magnetocapacitance
effect of LuFe$_2$O$_4$ at RT. According to our 
calculations, the two different CO structures, CO-I and
CO-III, are very close in energy. 
Given the room-temperature structure of hexagonal symmetry,
the CO-III state is higher in energy than the FE
CO-I state only by 20 meV/FU. Thus, at RT, there should
occur CFs associated with the interconversion between
the two different CO states, which should form different domains
separated by domain boundaries. 
Then, due to the large polarizability caused
by the CFs, the dielectric constant of LuFe$_2$O$_4$ should
be very large. 
Under an external magnetic field the Zeeman effect should
preferentially stabilize one of the two CO states because the two
states are most likely to 
have different total spin moments. Consequently, an external
magnetic field will reduce the extent of CF and hence
decrease the electron polarizability. 
The a.c. dielectric dispersion observed for LuFe$_2$O$_4$ can be understood
in terms of the dielectric response and the motion of the
FE domain boundaries between the CO-I and CO-III states \cite{Ikeda2005}.
At present we cannot answer the question of which CO state will be
preferentially stabilized by an external magnetic field, because only
the FM spin arrangement was considered in the present work.
Further studies are necessary to address this question.

We note that the spontaneous EP  of LuFe$_2$O$_4$
 is much greater than that of the recently discovered multiferroic
 materials (such as TbMnO$_3$ \cite{Kimura2003}),
 so LuFe$_2$O$_4$ represents a promising candidate for
 novel magnetoelectric devices.

We thank Dr. D. G. Mandrus, Dr. M. Angst and Prof. M. A. Subramanian
for useful discussion. This work was supported by the Office
of Basic Energy Sciences, Division of Materials Sciences,
U. S. Department of Energy, under Grant No. DE-FG02-86ER45259.

%\clearpage

%\clearpage


\begin{thebibliography}{99}


\bibitem{Auciello1998} O. Auciello {\it et al.}, Phys. Today
  {\bf 51}(7), 22 (1998); M. Dawber {\it et al.},
  Rev. Mod. Phys. {\bf 77}, 1083 (2005). 

\bibitem{Cohen1992} R. E. Cohen, Nature (London) {\bf 358}, 136 (1992).

\bibitem{Ikeda2005}N. Ikeda {\it et al.}, Nature (London) {\bf 436}, 1136 (2005).

\bibitem{Subramanian2006}M. A. Subramanian
{\it et al.}, Adv. Mater. {\bf 18}, 1737 (2006).

\bibitem{Isobe}M. Isobe {\it et al.},
Acta  Cryst. C{\bf 46}, 1917 (1990).

\bibitem{Tanaka}M Tanaka {\it et al.}, J. Phys. Soc. Jpn. {\bf 53}, 760 (1984).

\bibitem{LuFe2O4}Y. Yamada {\it et al.},
Phys. Rev. B
{\bf 62}, 12167 (2000); N. Ikeda, 
{\it et al.},
J. Phys. Soc. Jpn. {\bf 69}, 1526 (2000); J. Iida {\it et al.},
Physica B {\bf 155}, 307 (1989).

\bibitem{RFe2O4}K. Yoshii {\it et al.}, Physica B {\bf 378}, 585
  (2006); N. Ikeda, {\it et al.},
Ferroelectrics {\bf 314}, 41 (2005),
  and references therein.

\bibitem{Wright2001}J. P. Wright {\it et al.},
  Phys. Rev. Lett. {\bf 87}, 266401 (2001).

\bibitem{Yamada1997}Y. Yamada {\it et al.},
  J. Phys. Soc. Jpn. {\bf 66}, 3733 (1997).

\bibitem{PAW}P. E. Bl\"ochl, Phys. Rev. B {\bf 50}, 17953 (1994); G. Kresse and
  D. Joubert, {\it ibid}  {\bf 59}, 1758 (1999). 

\bibitem{VASP}G. Kresse and J. Furthm\"uller, Comput. Mater. Sci. {\bf
  6}, 15 (1996); Phys. Rev. B {\bf 54}, 11169 (1996).

\bibitem{Perdew1996}J. P. Perdew {\it et al.},
  Phys. Rev. Lett. {\bf 77}, 3865 (1996).

\bibitem{Dudarev1998}S. L. Dudarev {\it et al.},
Phys. Rev. B {\bf 57}, 1505 (1998).

\bibitem{Attfield}J. P. Attfield {\it et al.},
  Nature (London)  {\bf 396}, 655 (1998).

\bibitem{Leonov}I. Leonov {\it et al.},
  Phys. Rev. B {\bf 72}, 014407 (2005); J. Garc\'ia
  and G. Sub\'ias, {\it ibid} {\bf 74}, 176401 (2006); I. Leonov {\it
  et al.}, {\it ibid}  {\bf 74}, 176402 (2006). 

\bibitem{To_be_published}H. J. Xiang and M. -H. Whangbo, unpublished.

\bibitem{BVS} I. D. Brown and D. Altermatt, Acta Cryst. B {\bf 41}, 244 (1985).

\bibitem{largeU} 
  We first obtain the CO states using a large $U$ (e.g., $U_{eff}$ =
  7.61 eV) and then recalculate the CO states at a smaller $U$ (i.e., $U_{eff}$
  = 4.61 eV) using the converged densities with the larger $U$. 
  A large $U_{eff}$ was employed only to generate an initial
  electron density with a desired CO state.
  
\bibitem{Berry}R. D. King-Smith and D. Vanderbilt, Phys. Rev. B {\bf 47}, 1651
  (1993); R.Resta, Rev. Mod. Phys. {\bf 66}, 899 (1994).

\bibitem{Kimura2003}T. Kimura {\it et al.}, Nature (London) {\bf 426}, 55 (2003).


\end{thebibliography}
\end{document}